# Coupling graphene nanomechanical motion to a single-electron transistor

Gang Luo, Zhuo-Zhi Zhang, Guang-Wei Deng[*], Hai-Ou Li, Gang Cao, Ming Xiao, Guang-Can Guo, and Guo-Ping Guo[*]

Key Laboratory of Quantum Information, University of Science and Technology of China, Chinese Academy of Sciences, Hefei 230026, China.

Synergetic Innovation Center of Quantum Information and Quantum Physics, University of Science and Technology of China, Hefei, Anhui 230026, China.

*Correspondence to: gwdeng@ustc.edu.cn or gpguo@ustc.edu.cn

**Abstract:**

Graphene-based electromechanical resonators have attracted much interest recently because of the outstanding mechanical and electrical properties of graphene and their various applications. However, the coupling between mechanical motion and charge transport has not been explored in graphene. Here, we studied the mechanical properties of a suspended 50-nm-wide graphene nanoribbon, which also acts as a single-electron transistor (SET) at low temperature. Using the SET as a sensitive detector, we found that the resonance frequency could be tuned from 82 MHz to 100 MHz and the quality factor exceeded 30000. The strong charge-mechanical coupling was demonstrated by observing the SET induced ~140 kHz resonance frequency shifts and mechanical damping. We also found that the SET can enhance the nonlinearity of the resonator. Our SET-coupled graphene mechanical resonator could approach an ultra-sensitive mass resolution of $\sim 0.55 \times 10^{-21}$ g and a force sensitivity of $\sim 1.9 \times 10^{-19}$ N/(Hz)$^{1/2}$, and can be further improved. These properties indicate that our device is a good platform both for fundamental physical studies and potential applications.

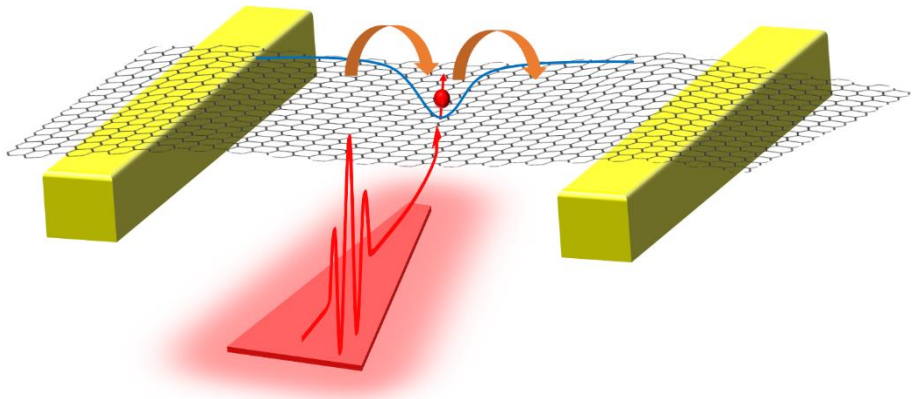

TOC

**Introduction:**

Because of its large Young's modulus (~1 TPa) and low mass, graphene is extremely promising for use in nanoelectromechanical systems (NEMS).[1-4] A graphene mechanical resonator is considered an ideal force[5] and mass sensor[6] because of its large resonance frequency and high quality factors, and it is also expected to exhibit quantum behavior at low temperature.[7-9] Conversely, because of its high electron mobility,[10] graphene has also been studied for etched[11] or gate-defined[12] quantum dots, which have been proposed as quantum bits.[13, 14] Suspended graphene has shown ultrahigh electron mobility,[15] offers a platform to observe the fractional quantum Hall effect,[16, 17] and it also performed well as quantum dots.[18-20] Thus, these properties open the way to study the interplay between mechanical and electronic degrees of freedom in graphene. Previous experiments have demonstrated ripple texturing of suspended graphene,[21] phonon softening of strained graphene,[22] a strain-induced zero-field quantum Hall effect,[23] and 300 Tesla pseudo-magnetic fields.[24] The relationships between strain and electrical resistance of suspended graphene have also been studied by AFM.[25, 26] Graphene has potential applications at room temperature: quantum Hall effect[27] and quantum dot behavior[28] at 300 K have been reported. Moreover, graphene nanoribbon is easy to scale up.[3, 29, 30] However, an experiment studying the coupling between a mechanical resonator and single-electron transistor (or quantum dot) in a graphene system is still lacking.

Many theoretical investigations have studied the interplay between mechanical motion and single-electron tunneling, where the mechanical motion is reported to affect the electron transport current[31] and current noise;[32] in turn, the back-action from the electron tunneling should cause frequency shifts[33] and damping[33, 34] in the mechanical modes. There are proposals to realize single-electron shuttles by mechanical motion[35, 36] and ground state cooling of the mechanical motion by quantum dots;[37] thus, further study could be performed in the quantum regime.[38, 39] Several pioneering experiments have demonstrated superconducting single-electron transistors as position detectors,[40-42] and strong coupling between single-electron tunneling and mechanical motions have been reported in carbon nanotubes.[43-46] In the case of graphene resonators, however, it is unknown whether the charge transport can be coupled to the mechanical motion.

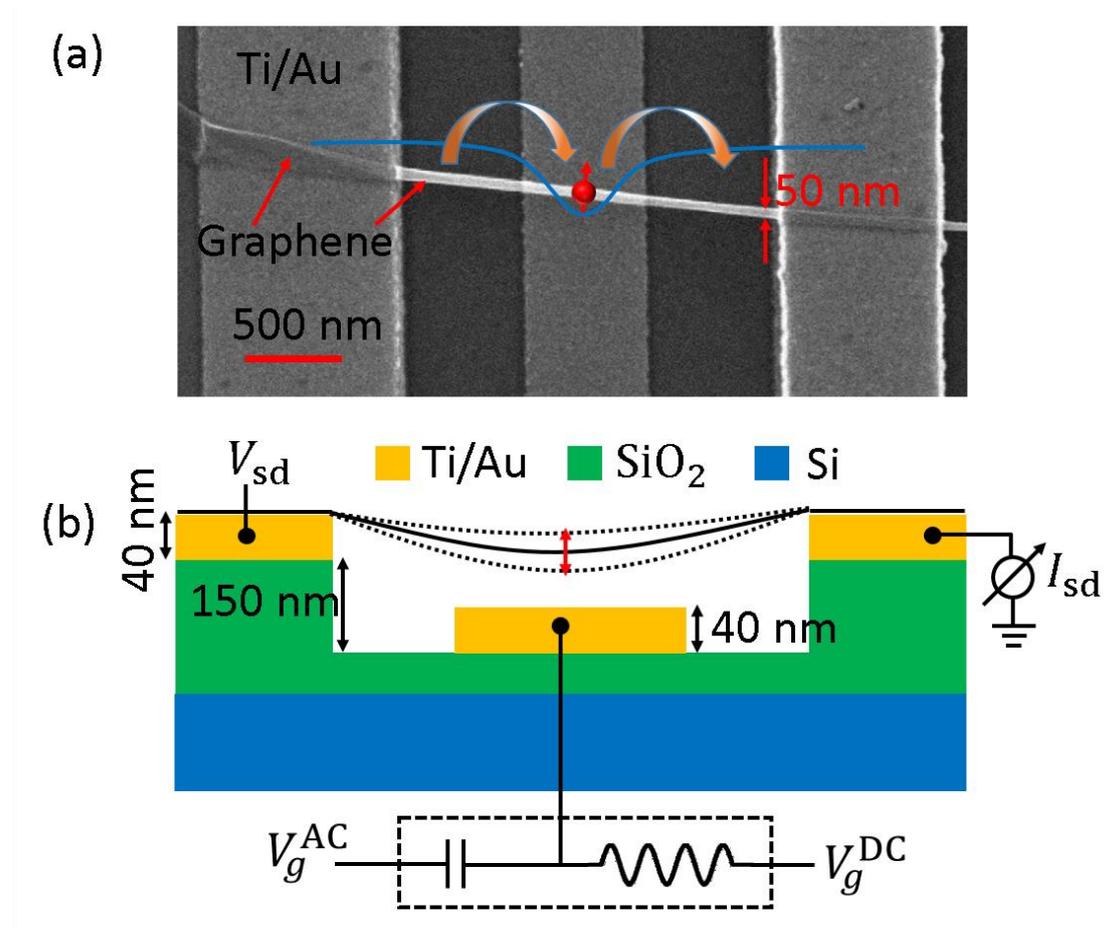

**Figure 1.** (a) Scanning electron microscopy figure of the nanoribbon-type graphene mechanical resonator. The doubly-clamped resonator was fabricated using mechanically exfoliated graphene. Two Ti/Au metal contacts were used as the source (*S*) and drain (*D*) of the quantum dot. A bottom Ti/Au gate beneath the graphene ribbon was used to apply DC and AC voltages, which can tune the chemical potential of the quantum dot, actuate the resonator, and change the mechanical tension. The blue curve shows the schematic potential, and the red dot indicates the quantum dot. (b) Schematic diagram of the sample and the measuring circuit. The substrate was undoped Si with 300 nm of $SiO_2$. A bias-T, shown in the dashed box, was used to combine DC and AC voltages applied to the bottom gate.

In this letter, we report the first experimental realization of coupling between a single-electron transistor and mechanical motion in a graphene nanoribbon. We find that the mechanical motion can be greatly affected by single-electron tunneling into and out of the graphene ribbon. The electron transport induced frequency shift is measured to be as large as 140 kHz, which is 56 times larger than the smallest resonance linewidth of our sample. We further observed a large reduction in the quality factor in the quantum dot region, which could be useful to study the nonlinear damping and cooling of graphene resonators.[47, 48] Moreover, the quantum dot affected the nonlinear dynamics of the resonator. These results show a strong coupling between electron transport and mechanical motion in graphene, which could be very useful in future high-frequency resonator measurements and for cooling the mechanical oscillations to the quantum regime. Graphene resonator has been demonstrated to interact with magnetic field recently,[49] and we expect more new physics when a quantum dot participates in this system in the future.

**Results:**

Our sample structure is shown in Figure 1, where a 50-nm-wide ($W$), 2-μm-length ($L$) graphene nanoribbon was suspended across two metal electrodes. A 5-layer (confirmed with AFM) graphene ribbon was transferred using an all-dry viscoelastic stamping technique,[50] without any further chemical etching process, which may harm the edge of the ribbon and leave chemical residues. A 40-nm Ti/Au (5 nm Ti and 35 nm Au) bottom electrode was deposited in a 150-nm deep (confirmed with AFM) silicon-oxide groove by e-beam lithography and a lift-off procedure, together with the two contact electrodes. Thus, the graphene mechanical resonator is expected to have a 150-nm distance from its bottom gate, ideally. We chose undoped Si with 300 nm of $SiO_2$ as the substrate because doped Si may absorb microwave power. However, this made it difficult to tune the chemical potential of the sample in a large range. The only way to control the potential in our sample is through the bottom gate.

The sample was mounted in a He-3 refrigerator with a base temperature of approximately 270 mK and at pressures below $10^{-6}$ torr. We actuated the suspended resonator by applying an AC field ($V_g^{AC}$) to the bottom gate, together with the DC voltages ($V_g^{DC}$), using a bias-T (Figure 1b). We applied a DC bias ($V_{sd}$) to the source contact and simply measured the DC current ($I_{sd}$) from the drain contact (Figure 1b). The ribbon was first annealed by the in-situ current annealing method (see the supplementary information). Previously, graphene nanomechanical resonators have been commonly analyzed by optical[1, 3] and mixing[2] techniques or by cavity-based capacitive readout methods.[8, 51] These methods require careful treatment of the parasitic capacitance and become difficult when the resonance frequency is as large as 1 GHz.[52] Quantum dots have been shown to detect mechanical motion with resonance frequencies up to 39 GHz in carbon nanotube devices,[53] and we expect a similar result in graphene systems for future high-frequency samples. Our 50-nm-wide graphene nanomechanical resonator also acts as a quantum dot, and we show its transport properties in Figure 2b, where the Coulomb blockade is clearly observed as a function of the gate voltage. The charging energy is estimated to be ~1 meV (see the supplementary information), and this kind of nanoribbon-type graphene quantum dot has been previously studied by several groups.[18, 19, 54]

Figure 2c shows a typical DC current response as a function of the driving frequency voltage. To obtain better resolution, the data were acquired using the differential of the frequency axis. Two resonance modes can be observed by the transport measurement. The first mode is "W" type, and the second one could not be resolved at small gate voltage. (b) Transport property of the sample as a function of the gate voltage, which shows clear Coulomb peak structures. The bias voltage $V_{sd}$ is fixed at 20 µV for this curve. (c) Fitting the current as a function of the driving frequency. The data were obtained at -70 dBm, as the average of 10 measurements. (d) Zoom of the dashed black box in panel a. The related resonance frequency goes up and down, corresponding to the Coulomb peak and blockade regime. This diagram means that single-electron tunneling modulates the resonance frequency of the mechanical resonator, indicating that the two systems are coupled.

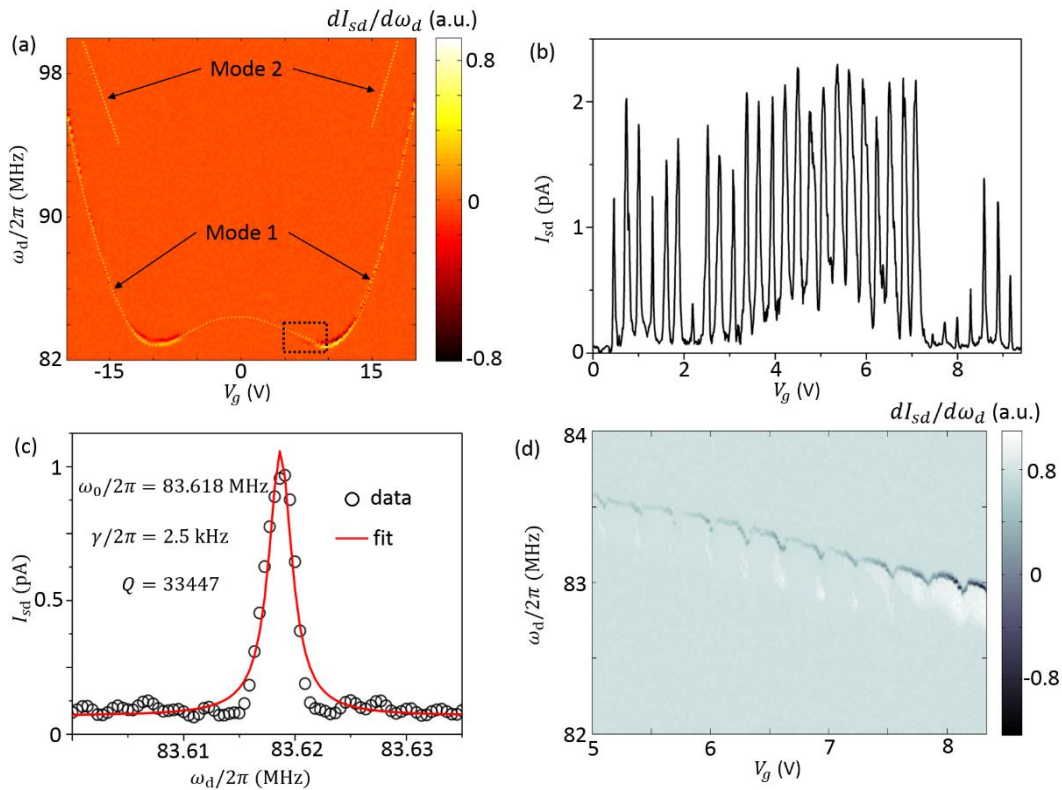

**Figure 2.** (a) Current response as a function of the driving frequency and the gate

near its resonance center. Here the DC and AC signal applied to the bottom gate lead to a mean electrostatic force on the resonator,[45]

$$F = \frac{1}{2}\frac{\partial C_g}{\partial z}(V_g^{DC} - V_{dot} + V_g^{AC})^2, \qquad (1)$$

where $C_g$ is the capacitance between the bottom gate and the resonator, and $V_{dot}$ is the potential of the graphene ribbon and depends on the occupation of the charge number;[43, 44, 55] through the capacitance, the DC voltage $V_g^{DC}$ can affect the tension of the resonator, which may change the resonance frequency of the resonator (see Figure 2a). In addition, the gate voltage also controls the electrochemical potential of the quantum dot, as we have discussed previously. The change in $z$ is with respect to the displacement vibration $\delta z(\omega, t) = A(\omega)\cos(\omega t + \phi)$. Here, $\omega = 2\pi f$ is the circle frequency of the driving AC signal and $\phi$ comes from the phase difference between the driving field ($V_g^{AC} = V_g^{RF}\cos(2\pi f t)$) and the vibration. When the AC driving frequency $f$ approaches the resonance frequency $f_0 = \omega_0/2\pi$, the mechanical vibration is effectively actuated and modulates the electrochemical potential of the quantum dot, resulting in a change in the conductance; thus, a current peak/dip (Figure 2c) can be observed. The current changes with time as $I_{SD}(t) =$

$$\sum_n \frac{1}{n!} \frac{d^n I_{SD}^{DC}(V_g)}{dV_g^n} \left[ \frac{V_g^{DC}}{C_g} \frac{\partial C_g}{\partial z} \delta z(\omega, t) \right]^n, \text{ resulting in a measured current change of}^{45}$$

$$\Delta I_{sd}(t) \approx \frac{1}{4} \frac{d^2 I_{SD}^{DC}(V_g)}{dV_g^2} \left[ \frac{V_g^{DC}}{C_g} \frac{\partial C_g}{\partial z} A(\omega) \right]^2, \quad (2)$$

where $A(\omega) = \dfrac{\frac{\partial C_g}{\partial z} V_g^{DC} \delta V_g^{RF}}{m_{\text{eff}}} \times \dfrac{1}{\sqrt{(2\omega_0(\omega_0-\omega))^2 + \frac{\omega_0^4}{Q^2}}}$ (3)

is the amplitude of the mechanical resonator without considering of the nonlinearity at very low driving powers. In Figure 2c, we show a typical fitting of the current change as a function of the driving frequency and obtain a line width $\gamma = 2.5$ kHz, which corresponds to a quality factor $Q = \omega_0/\gamma = 33447$ and yields an energy relaxation time of 400 μs. The frequency-Q product is usually considered as the figure of merit when comparing different mechanical resonator systems. We reached a frequency-Q product of $fQ \sim 3 \times 10^{12}$ Hz, which is comparable with that of carbon nanotube devices.[56] We estimated the effective mass of the resonator to be $m_{\text{eff}} \sim 1.85 \times 10^{-19}$ kg for the $N = 5$-layer graphene nanoribbon, where $m_{\text{eff}} = 0.5 N \rho L W$ and $\rho = 7.4 \times 10^{-19}$ kg/μm². In the situation with low driving power, the force can be simply treated as $F = k\delta z$, where $k$ is the spring constant. Simply, the resonance frequency can be obtained as $f_0 = \frac{1}{2\pi}\sqrt{\frac{k}{m_{\text{eff}}}}$, from which we extract $k \sim 0.05$ N/m.

We further study the interaction between the quantum dot and the mechanical resonator in Figure 3. Compared with Figure 2c, Figure 3 was measured with larger driving power (-50 dBm) and better resolution, but a smaller quality factor. We find that $\omega_0$, $\gamma$, and $Q$ (Figure 3b,c,d) all oscillate as a function of the gate voltage, together with the Coulomb oscillations shown in Figure 3a. The resonance frequency shift is caused by the back action force induced by the fluctuation of electron transport on the resonator. We found that the largest frequency shift in our sample is approximately $\Delta\omega_0 = 140$ kHz $\times 2\pi$, which is 56 times larger than the smallest resonance linewidth we have measured, indicating a strong coupling of the graphene mechanical motion and the single-electron tunneling. Even compared with the same measurement shown in Figure 3c, this frequency shift is 6 times larger than the largest linewidth around the Coulomb peak. From the blockade to the Coulomb peak, the electron-phonon coupling induced the mechanical damping, increasing the linewidth from 13 kHz to 26 kHz and accordingly reducing the quality factor from 6000 to 3000. This result means that charge transport plays an important role in the mechanical dissipation mechanism at low temperature. The large frequency shifts and damping rates caused by the Coulomb peaks indicate a strong coupling between graphene mechanical motion and single-electron tunneling.[43-45] This kind of strong coupling could be very useful to cool graphene mechanical resonators to the ground state and for further applications.

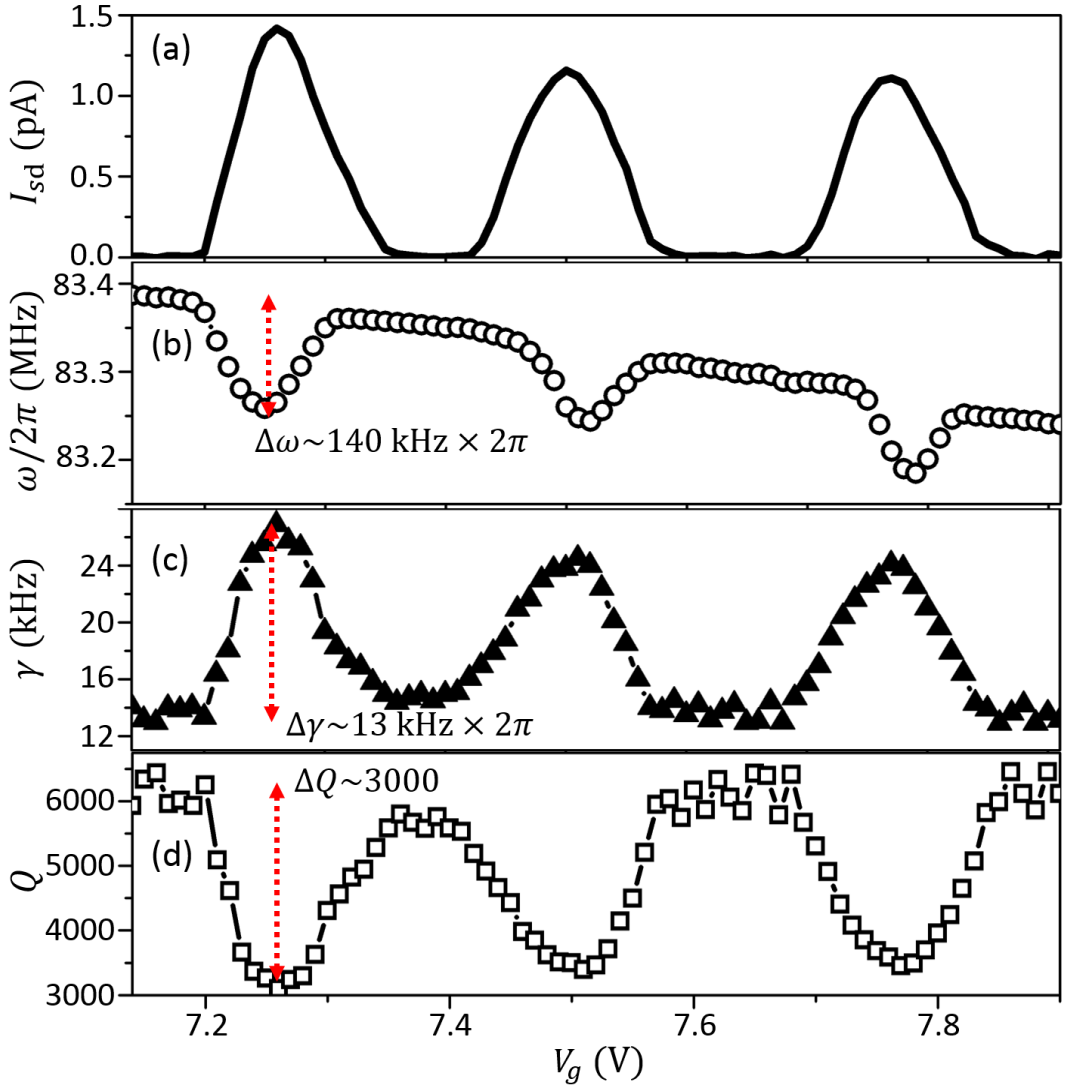

**Figure 3.** (a) Current as a function of gate voltage, showing three Coulomb peaks. (b-d) The corresponding frequency response (b), dissipation rate (c) and quality factor (d).

With large driving power, the resonator enters the nonlinear regime, which is a result

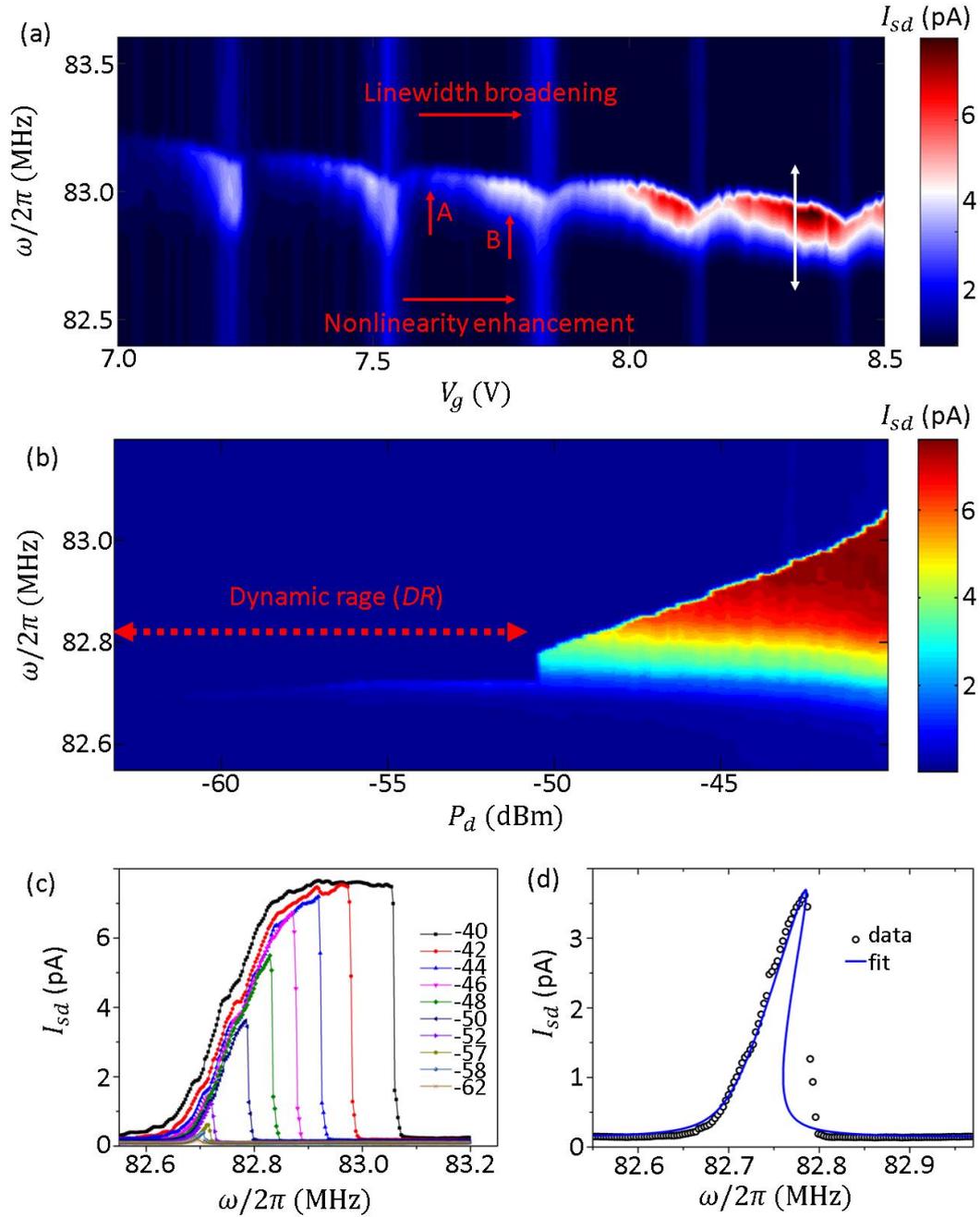

**Figure 4.** Nonlinear dynamics. (a) Transport property as a function of the driving frequency and the gate voltage at large driving power (-48 dBm), where the nonlinear response is clearly observed and enhanced by the quantum dot. (b) Current as a function of the driving power and frequency in the Coulomb blockade region (white arrow in panel a), where the background current is zero. Linewidth broadening can be clearly observed when the power is larger than -50 dBm. (c) Transport current as a function of the driving frequency, for various powers. The units are dBm. (d) Nonlinear fit of the measured data; the fitted Duffing coefficient in this figure is $\alpha = 2 \times 10^{16}$ kg/m$^2$s$^2$

and obtained from point A shown in panel a ($\alpha = 5 \times 10^{16}$ kg/m²s² for point B).

of the resonator being two dimensional. Without the effect from the quantum dot, we consider the Duffing nonlinearity term $\alpha(\delta z)^3$ and high order nonlinear damping term $\eta(\delta z)^2 \frac{dz}{dt}$, the Newton equation for a nonlinear resonator can then be described as:[47,56]

$$F = m_{eff}\frac{d^2\delta z}{dt^2} + k\delta z + \gamma\frac{d\delta z}{dt} + \alpha(\delta z)^3 + \eta(\delta z)^2\frac{d\delta z}{dt}. \qquad (4)$$

Regarding the nonlinearity induced by the SET, we describe the SET force by:[55]

$$F_{SET} = -k_{SET}\delta z - \beta_{SET}(\delta z)^2 - \alpha_{SET}(\delta z)^3, \qquad (5)$$

where the SET induced nonlinearity $\alpha_{SET}$ can be simply renormalized to the total Duffing term.[55] And $\delta z$ is defined as the small vibrations around a static equilibrium $z_0$ (a fixed reference point), which depends on the DC voltage $V_g^{DC}$.

Figure 4 shows the nonlinear dynamics of the sample, where the nonlinear induced linewidth broadening becomes obvious when the driving power is larger than -50 dBm (Figure 4b). This gives a 20 dB dynamic range (*DR*), which is very useful for mass sensing. The mass resolution was previously given as $\delta m = \frac{m_{eff}}{Q} \cdot 10^{-DR/20}$,[57] from which we obtained a mass resolution of $\sim 0.55 \times 10^{-21}$ g (or 0.55 zeptograms). The measured transport current was found to be saturated as a function of driving frequency at very high driving power (Figure 4c). Neglecting the high order damping term, we fit the Duffing coefficient to be $\alpha \sim 2 \times 10^{16}$ kg/m²s² in a Coulomb blockade region (Figure 4d), which is comparable with a previous report.[56] We also found that quantum dots could enhance the Duffing nonlinearity (Figure 4a), where the Duffing coefficient in the quantum dot region increased to $\alpha \sim 5 \times 10^{16}$ kg/m²s². This single-electron tunneling enhancing Duffing nonlinearity has been reported in a carbon nanotube system;[55] however, it has not been observed in graphene. Exploring the onset of nonlinearity in a two-dimensional material is quite interesting and important, and may bring new understanding of high-order nonlinear physics.

The QD-coupled mechanical resonator has a small *DR*; however, the small mass and high quality factor contribute to a high sensitivity (at zg level) for the mass resolution.

A wider device will have higher *DR*; however, it will have a larger mass (see supplementary information). Because the device has a large frequency-*Q* product, we expect a large force sensitivity $F_{min} = (4k_b Tk/\omega Q)^{1/2}$,[58] where $k_b$ is the Boltzmann constant and $T = 270$ mK is the temperature. We calculate the force sensitivity to be $F_{min} \sim 1.9 \times 10^{-19}$ N/(Hz)$^{1/2}$ (or 19 aN/(Hz)$^{1/2}$) for the current setup. In future work, by cooling down the sample to lower temperature and enhancing the quality, this sensitivity could be further optimized.

In conclusion, we have realized the first coupling between the mechanical mode and single-electron tunneling in a graphene device. We find that the 50-nm-wide graphene resonator shows a high quality factor of $> 3 \times 10^4$ and resonance frequency as high as 100 MHz. Therefore, the resonator could act as a good mass and force sensor. The resonator also acts as a quantum dot at low temperature, and single-electron tunneling causes a $>$ 100-kHz frequency shift of the resonator. Our device also offers a platform for the study of nonlinear physics with respect to quantum dots. Further investigations may use this device for detecting ultra-high-frequency resonator and cooling the mechanical modes to the quantum regime.

**Acknowledgment**

This work was supported by the National Key R & D Program (Grant No.2016YFA0301700), the Strategic Priority Research Program of the CAS (Grant No. XDB01030000), the National Natural Science Foundation (Grant Nos. 11304301, 11575172, 61306150, and 91421303), and the Fundamental Research Fund for the Central Universities.

**Competing financial interest:**

The authors declare that they have no competing financial interests.

Supporting Information Available:

Further information concerning the placement method for graphene, sample preparation, measurement setup, Coulomb diamond, Duffing coefficient fitting, and comparison with another sample is available. This material is available free of charge via the Internet at http://pubs.acs.org.